# GameControllerizer: Middleware to Program Inputs for Augmenting Digital Games

Kazutaka Kurihara*, Nobuhiro Doi**

* Tsuda University
** Kurusugawa Computer Inc.

**Abstract.** This study proposes middleware, GameControllerizer, that allows users to combine the processes of Internet of Things (IoT) devices, Web services, and applications of Artificial Intelligence (AI), and to convert them into game control operations to augment existing digital games. The system facilitates easy trial-and-error development of new forms of entertainment and the configuration of gamification by enabling the use of diverse devices and sources of information as inputs to games. GameControllerizer consists of a visual programming element that uses the Node-RED tool to allow users to program easily to convert diverse formats of information into inputs to games, and contains a game input emulation element whereby hardware- and software-based emulation generates inputs for gaming devices. Evidence of the usefulness of the system was provided by a performance assessment and the proposal of a variety of use cases.

**Keywords:** Gamification, Toolification of Games, Input Emulation, Node-RED, IoT.

## 1 Introduction

The advent of the Internet of Things (IoT); whereby a diversity of everyday objects connected to the Internet can behave intelligently; various Web services that can be used by end-user programmers through publicly available APIs; and Artificial Intelligence (AI) technologies such as machine learning, which have rapidly improved and become more accessible in recent years, have created possibilities for building information systems that can enrich our everyday lives. The abilities of the general public to create such information systems as a hobby have matured with the popularization of maker culture, and a vast array of examples can now be seen at hackathons, exhibitions, and on video-sharing sites.

The focus of our research is to explore support technologies that facilitate easier utilization and augmentation of existing digital games, which are a commonly used medium for such composite information system development owing to their high level of recognition by the general public [1][2][3]. The augmentation of digital games can provide new entertainment value and facilitate the achievement of non-game purposes, such as the implementation of social benefits. The latter characteristic refers to the "toolification of games," a peripheral concept in gamification proposed by Kurihara [4]. Thus far, however, the development of augmentation systems that reuse digital games has been hindered due to issues concerning intellectual property rights and difficulties in obtaining source codes.

In the above context, this paper proposes an external augmentation architecture for digital games. This information system architecture works with digital games in their original state but augments them by estimating the internal state of the game using pattern recognition technology to detect and recognize sound and image outputs as well as performing arbitrary information processes, the results of which are converted into inputs that provide feedback to the game (Fig. 1). An important characteristic of this architecture is that it enables the use of games without the need to change them, thus solving the intellectual property rights issue and the difficulties in obtaining source codes. While examples of the development and employment of this kind of architecture in an ad-hoc fashion are available, the contribution of this study lies in formulating the system structure, dividing it into components, and developing and offering uses of the respective support technologies.

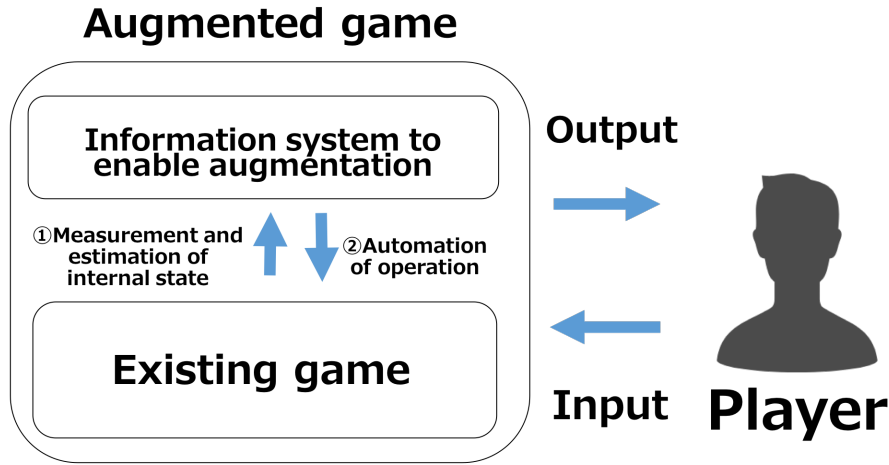

**Fig. 1.** External augmentation architecture for digital games.

Thus far, developer support for the element "Measurement and estimation of internal state" in Fig. 1 has been provided through technologies such as Sikuli [5], which handles images, and Picognizer [6], which handles electronic sound. On the contrary, little developer support is available for the "Automation of operations," i.e., the element that performs conversion and connects to the input of the game controller to control the internal state of the game as intended by the augmentation information system. Thus far, when development is undertaken, it has often been on an ad-hoc basis, using such methods as emulating the keyboard or mouse with a UI automation library and/or directly altering the electronic circuit of the game controller to produce an input signal.

To simplify this common and repeated development labor, this paper proposes GameControllerizer, middleware for transforming diverse devices and sources of information into inputs for augmenting existing digital games. By enabling the use of diverse devices and sources of information as inputs to games, the system facilitates easy trial-and-error development of new forms of entertainment and gamification.

GameControllerizer consists of (1) a game input emulation element that emulates inputs for a number of digital game platforms via software (S/W) and hardware (H/W) elements; and (2) a visual programming element whereby the end-user programmer communicates with diverse devices and sources of information, and expresses the operation of input transmission for the game input emulation element. The latter uses the Node-RED tool [7] for a visually user-friendly programming experience. The time series information for the input signal to the game is written in DSL4GC (Domain-specific Language for Game Control), a simple and highly readable language written in one-dimensional character strings. The GameControllerizer software has been released as open source and the hardware is scheduled for distribution.

This paper first provides an overview of the relevant literature in Section 2. The implementation of GameControllerizer is described in Section 3. Subsequently, evidence of the utility of the system is provided by its assessment in terms of delay performance as a fundamental factor in Section 4, and by proposing a variety of use cases in Section 5. A discussion of outstanding challenges and directions for future research are presented in Section 6, followed by the conclusion in Section 7.

## 2   Related Work

### 2.1   Input Device Emulation

This study deals with programming technologies that use input device emulation to make a given input appear to behave like a real person. Research in the field of GUI automation is progressing on this theme. The Go libraries Robotgo [11] and Automator [9], a standard feature in Mac OS, and similar offerings enable the automation of GUI operations in text, vision, and programming through demonstration formats. Sikuli [5] is a GUI automation programming environment where screenshots of objects, such as icons, are pasted directly using Python programming code to instruct the system to regard the image as the target. Kato et al.'s Picode [10] augments Sikuli to use snapshot data of such things as human movement or the position of a robot as detection targets.

Moreover, hardware products, such as Phantom-S [12], that can emulate gamepad inputs are available, but have problems related to little programmability of input behaviors for games.

In addition to the above, the automation of gameplay is a common theme in AI research. Fujii et al. [13], for instance, conducted an assessment of how human-like the automated operation of the game Super Mario Bros. was. Gym Retro [14] is a publicly available game emulator for AI research that enables the programming of automated operations.

By using Node-RED as programming environment and employing both hardware- and software-based game controller emulation, this study differs from past research in that it specializes in the provision of two types of support—programming support and hardware construction support—to convert diverse sources of information from the Internet into game inputs. With regard to research related to game input augmentation, a notable innovation is Makey Makey [22], a hardware emulator that detects touch gestures for everyday objects and converts them into keyboard and mouse inputs. An important characteristic of this study is that it allows for the use of diverse input devices, including Makey Makey, not just in terms of simple key mapping, but also by facilitating the programming of more complex input behaviors for games.

**2.2  Examples of Reuse of Games**

The augmentation of digital games is a common theme in the maker movement and at hackathons. For instance, at "Asobi no Re-hack" ("Re-hacking Play"), a hackathon event where contestants reimagine existing kinds of "play," a project based on Tetris won the grand prize [1]. Life-size Katamari [2] is a personal project that involves a giant trackball as controller for the game Katamari Damacy. In Twitch Plays Pokemon [3], a large and unspecified number of viewers control the game by providing inputs to the game controller in text form via chat. The positive implications of this type of game design, where an unspecified and large number of players play a single game, have inspired follow-up research [15][16].

Furthermore, independent experiments to add new entertainment value to dancing games, such as Dance Dance Revolution [23], by incorporating hand movements that the system cannot detect (Kurihara calls it a "customized game" [4]), have grown in popularity, and can be easily found on video-sharing websites. To respond to growing interest in the reuse of games, as illustrated above, this study develops middleware that can use various sources of information as inputs to games through ad-hoc and easy-to-use programming. The target users are expected to be teams of designers and engineers who are likely to participate in events such as hackathons.

## 3  GameControllerizer

Fig. 2 shows the structure of GameControllerizer, the proposed game augmentation development support middleware. Subsequent sub-sections describe the structure and use of each element.

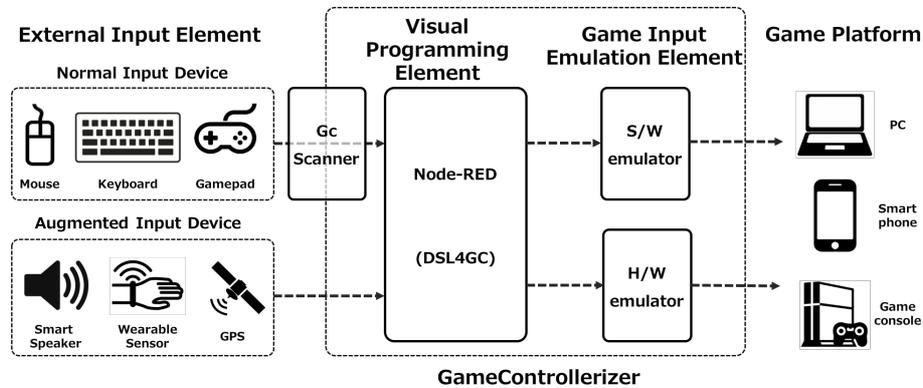

**Fig. 2**. Structure of GameControllerizer.

### 3.1 External Input Element

The external input element sits outside GameControllerizer and handles inputs from the user. Two types of devices constitute this element: devices that are generally used to provide inputs to games and augmented input devices, which are not normally used in this way. The former type consists of gamepads, keyboards, and mice that are made connectable to the visual programming element using GcScanner. GcScanner is a multiplatform helper tool that operates on a Web browser, converts user input into DSL4GC, and transmits it via WebSocket.

Augmented input devices are input devices and information sources that are not generally considered input devices for normal games. These can include smartphones, smart speakers, wearable sensors, GPS, optical cameras, weather forecasting Web APIs—things that may not even have a direct relationship to gaming. Information from these sources is transmitted to the visual programming element using protocols, such as HTTP and MQTT, thus enabling them to be used as input devices for games.

### 3.2 Visual Programming Element

The visual programming element is a programming environment that prescribes the logic used for transforming the information relayed by the external input element into inputs for the game. It employs Node-RED.

Node-RED is a framework for receiving, converting, and transmitting messages on a server. As all processes are defined by process blocks (nodes) on a GUI, it is possible to apply them to the input data by simply connecting them together. Accordingly, it has low learning cost, which renders it suitable for the type of rapid prototyping assumed in this study.

**DSL4GC.** In GameControllerizer, game control information is expressed in the abstract format DSL4GC (Domain-specific Language for Game Control). Appropriately changing this information and re-exporting it in Node-RED makes it possible to map it to inputs for the games. Fig. 3 shows the grammar of the DSL4GC. An operation where the direction keys and analog control stick are in the neutral position and button one is pressed for two frames (2/60 s), for example, would be expressed in JSON as follows:

*{"dpad":5, "btn":[1], "dur":2, "ang":[0,0,0,0]}*

As the above example shows, the DSL4GC has a high level of readability. Refer to [8] for detailed specifications of the DSL4GC. A joystick/mouse/keyboard format for common control messages for devices connected to computers that conform to HID specifications is available [17], but while highly extensible, it is extremely complex, rendering it over-specified for the simple control and augmentation of general games. As such, we define this domain-specific language to easily handle the necessary and sufficient control information. As the JSON format is compatible with existing definitions while remaining extensible, the language can be extended as required by future applications.

```
gc_sentence = Array[gc_word]
gc_word = gc_gamepad_word | gc_mouse_word | gc_keyboard_word
gc_gamepad_word = {"dpad":Int, "btn":Array[Int], "ang":Array[Int], "dur":Int}
gc_mouse_word = {"btn":Array[Int], "mov":Array[Int], "dur":Int}
gc_keyboard_word = {"key":Array[String], "mod":Array[Int], "dur":Int}
```

**Fig. 3**. Grammar of DSL4GC.

**Dedicated auxiliary nodes in the visual programming element.** In the GameControllerizer environment, the DSL4GC represents game control information and is generated either by a normal device via the GcScanner or by the visual programming element using input from the augmentation device as a trigger, which is then transmitted through Node-RED. Dedicated auxiliary nodes were developed in order to program this process more easily. These are the remap-button node, remap-dpad node, and remap-ang node, each of which corresponds to the buttons, direction keys, or analog stick. There are also virtual device nodes (three types: virtual gamepad, virtual keyboard, and virtual mouse) that generate the DSL4GC in Node-RED. Finally, an H/W emulator node exports the DSL4GC to the game input emulation element.

### 3.3 Game Input Emulation Element

The game input emulation element is the output of GameControllerizer. It provides input to the platform that operates the game. It converts and transmits the DSL4GC in the control signal form specified by each platform. GameControllerizer is equipped with two forms of input emulation.

**H/W Emulator.** The H/W module electrically emulates the basic operations of the normal devices discussed above (Fig. 4, left). The H/W emulator is equipped with inputs to the game platform via a USB and a unique connector as well as a UART control input (Fig. 5). This makes it possible to electronically control the keyboard, mouse, or gamepad via the DSL4GC. It was implemented on an mbed development environment [18], compatible ARM microcontroller, and HID device emulation libraries [19][20].

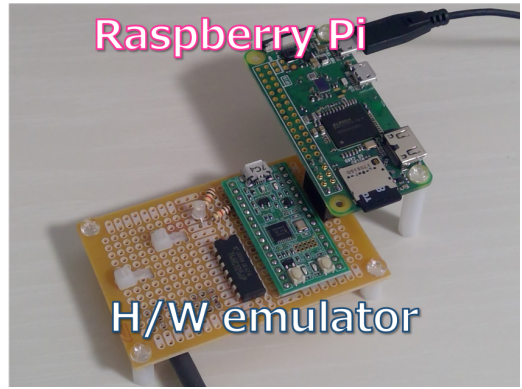

**Fig. 4.** (Left) H/W emulator and (right) Raspberry Pi Zero W, which runs the visual programming element.

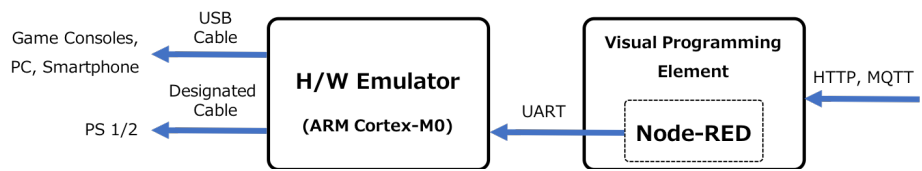

**Fig. 5.** Structure and method of communication of the H/W emulator.

Modifying the firmware of the ARM microcontroller (which handles the core processing for the H/W module) makes it behave as an input device for PC-based, smartphone-based, or PlayStation 1/2 games. As USB HID-specified devices can be connected to major gaming platforms, such as the PC, Playstation3/4, XBOX/XBOX-360, and Nintendo Switch, either directly or via a commercially available converter, the H/W structure described above can control majority of gaming platforms.

Further, the H/W module is designed to easily connect to Raspberry Pi Zero W (Fig. 4). Running the visual programming environment described above on Raspberry Pi Zero W makes it possible to implement all features of the H/W emulation-type GameControllerizer with this set alone. This means that developers can augment existing games by simply connecting the set to a network.

**S/W Emulator.** The S/W emulator is a program that emulates the basic operations of an input device. It receives the DSL4GC via the MQTT, interprets it, and emulates the operations of the device accordingly (Fig. 6). It was implemented with the Go device emulation library RobotGo [11], thus making it multi-platform compatible. At present, only keyboard/mouse emulation is possible. A device emulation method to mimic the behavior of Bluetooth gamepads is currently being developed.

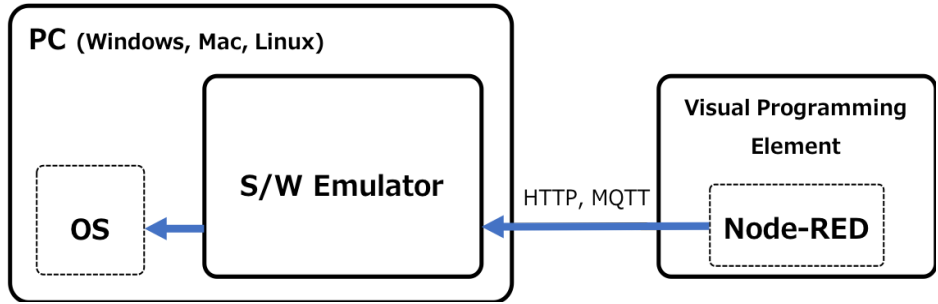

**Fig. 6.** Structure and method of communication of the S/W emulator.

### 3.4 Use Example of GameControllerizer

Fig. 7 shows the flow for executing a control whereby an HTTP GET request is received and a gamepad button is pressed. Fig. 8 shows the settings screen for a dedicated auxiliary node (virtual gamepad node). An augmented input device can be configured for use as a game input simply by using this GUI to define its settings. GET requests can be made in various ways, such as by PC browser access or sending events from programmable IoT devices such as the Sony MESH.

Alternatively, by replacing the virtual gamepad node with the function node, which is a standard feature of Node-RED, the following expression sends the "fireball" (i.e., *Hadouken*) move from the game Streetfighter II (moving the direction key down, lower right, right, and pressing Button 1):

```
msg.payload = [
  {"dpad":2, "btn":[], "dur":2, "ang":[0,0,0,0]},
  {"dpad":3, "btn":[], "dur":2, "ang":[0,0,0,0]},
  {"dpad":6, "btn":[1], "dur":2, "ang":[0,0,0,0]}
];
return msg;
```

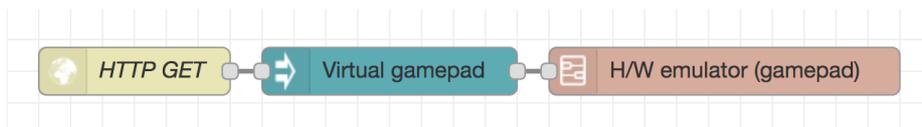

**Fig. 7.** Visual programming example for pressing a gamepad button after receiving an HTTP GET request.

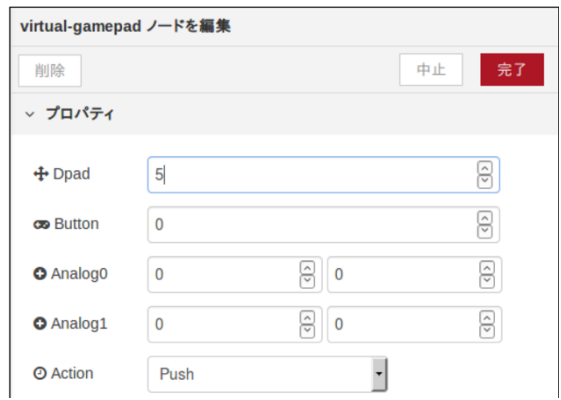

**Fig. 8.** GUI settings screen for virtual gamepad nodes.

## 4   Assessment of Basic Performance

This section discusses the method for and results of the assessment of the basic performance of GameControllerizer. Large transmission or processing delays in the middleware negatively impact gaming experience. Acceptable delay thresholds have been discussed in detail in research on the recently popular field of online gaming [21][24][25]. Table 1 lists the conclusions of [21]. Specifically, games comparatively sensitive to delay, such as first-person shooter (FPS) games, have a delay threshold of 100 ms, whereas the threshold for games comparatively insensitive, such as real-time strategy (RTS) games, is 1000 ms. We performed measurements and assessed the middleware in terms of whether it satisfied the delay thresholds for gaming. Note that although we refer to data for online games, the primary targets of the proposed system are standalone games, which can be played with minimal Internet access; thus, we can exclude possible Internet delays outside the local area network (LAN). In this assessment, the delay times used were the average values of 100 measurements of the same route.

**Table 1.** Game types and delay thresholds (from [21]).

| Group | Model | Perspective | Example Genres | Sensitivity | Delay Thresholds |
|---|---|---|---|---|---|
| 1 | Avator | First Person | FPS, Racing | High | 100 ms |
| 2 | Avator | Third Person | Sports, RPG | Medium | 500 ms |
| 3 | Omnipresent | Varies | RTS, Sim | Low | 1000 ms |

### 4.1 Method of Assessment and Results

Two types of assessment, stand-alone and overall, were performed. The stand-alone assessment measured delay within the proposed system, and the overall assessment measured overall delay when using the standard form of the proposed system in its expected use environment—connected to a standard household Wi-Fi LAN and no Internet access.

**Stand-alone assessment.** In the stand-alone assessment, the time taken to receive, process, and transmit the DSL4GC, which expresses singular-button input, via the MQTT was measured (Fig. 9). Three types of devices were used for the external input element, each of which was combined with all usable dedicated auxiliary nodes (for each device, respectively) created in the visual programming element. The combination of the three device types with the H/W and S/W emulation in the game input emulation element resulted in a total of six assessment conditions. Table 2 lists the results of the assessment.

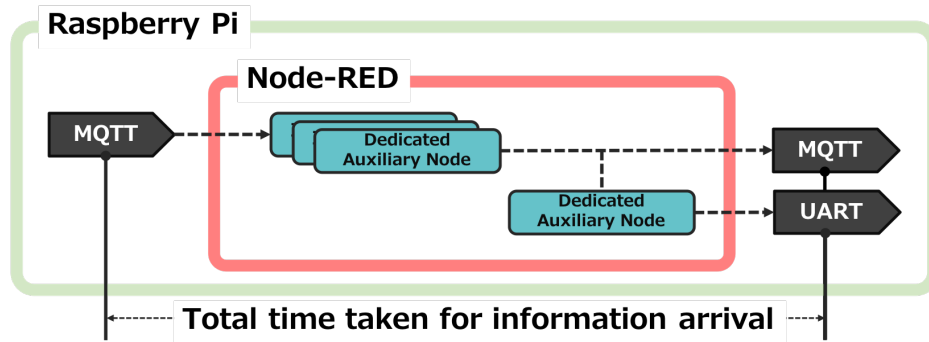

**Fig. 9.** Measurement environment for stand-alone assessment.

**Overall assessment.** In the overall assessment, the delay from scanning a single mouse input at 60 fps with GcScanner and transmitting it via Wi-Fi to the UART output of the H/W emulator element was measured (Fig. 10). To use all usable dedicated auxiliary nodes created in the in the visual programming element, the function, remap-button, remap-ang, remap-dpad, and HwEmulator gamepad nodes were connected. The average delay was 44 ms.

Table 2. Results of stand-alone assessment.

| Input | Processed Nodes | Emulation | Average Delay |
|---|---|---|---|
| Gamepad | remap-button<br>remap-ang<br>remap-dpad | S/W | 15 ms |
| Gamepad | remap-button<br>remap-ang<br>remap-dpad<br>H/W Emulator | H/W | 16 ms |
| Mouse | (none) | S/W | 9 ms |
| Mouse | H/W Emulator | H/W | 14 ms |
| Keyboard | (none) | S/W | 9 ms |
| Keyboard | H/W Emulator | H/W | 9 ms |

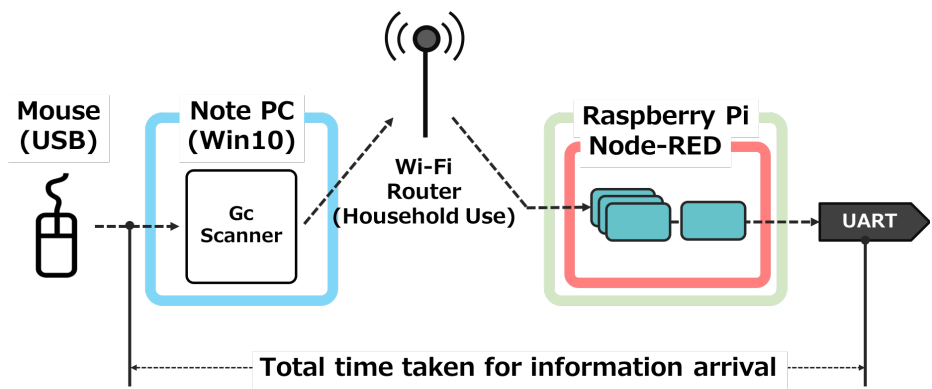

Fig. 10. Measurement environment for overall assessment.

### 4.2 Implications

For the stand-alone assessment, the largest increase in delay when using Node-RED and dedicated process modules as middleware was 16 ms. In light of the values in Table 1, this value did not significantly affect user experience in games where timing is critical.

Further, the delay for the overall assessment was kept to around 50 ms, with no recorded delay exceeding 100 ms. In this environment, delays were within the acceptable range for all of the three types of games in Table 1.

It is worth noting that the overall assessment included Wi-Fi delay, which needs to be considered in practice. The ping values between machines returned an average time of 7 ms. As the measurement environment was favorable in terms of obstacles and number of connected devices, the delay was kept at this level. Network delays can, however, become an overriding problem in cases where users do not have exclusive use of the network, or when controlling the game via an external MQTT server or APIs through the Internet. One way to address this issue is to host all environments (Node-RED, GcScanner) on a single computer.

## 5 Application Scenarios

This section presents feasible scenarios for the application of the proposed technique. Refer to the demonstration video for review to see working examples.

### 5.1 Augmentation of Input Devices

The proposed method enables the use of devices and sources of information that are not normally used as input devices to games. This feature contributes to discovering new entertainment value in existing games and enhancing accessibility for handicapped people to the games as well. For instance, Smartphone gestures, Bluetooth smartphone shutter devices, Sony MESH, physical movement while wearing a smartwatch, audio input from smart speakers, vision-based object tracking, and exchange fluctuations from an exchange API can be used as game inputs. Further, the system makes it possible to distribute the operation of a game over a small, or large and unspecified number of players, similar to the case of Twitch Plays Pokémon.

### 5.2 Configuring the Toolification of Games

As discussed above, the ability to use various devices and sources of information as game inputs makes possible the toolification of games, which is a method of gamification using existing games. Detecting the rhythm of a good shoulder massage with a sensor, such as a smartwatch, and using it as an input to a music game may be used to provide an effective massage or promote interpersonal communication, for instance. Alternatively, detecting shoe-shining motion or sweeping/wiping movement with an accelerometer and connecting it to a fun and exciting action game as an input to simple, repeated button-pressing operation may help sustain motivation at work.

### 5.3 Input Augmentation Co-existing with Existing Input Methods

The proposed method makes it possible to retain the standard way of playing by connecting a gamepad to the external input element while also enabling the insertion of auxiliary inputs as required. An example is using a hot key or voice input in place of a command that is particularly difficult to input. Moreover, even when handling various input devices, as discussed in Section 5.1, by limiting their use to specific in-game situations and using the standard inputs at other times, opportunities for the use of this system can be enhanced.

### 5.4 Handling Multiple Games

The proposed method enables the exploration of new forms of entertainment. In case of input from a player, for example, implementing simultaneous or selective inputs to multiple, independently running games can enable a fighting game, which is original-ly one-on-one, to be modified so that multiple enemies can be fought at once, or for a player to play both Tetris and Puyo Puyo simultaneously.

### 5.5 Creating Closed Loops of External Game Augmentation Architecture

By estimating the internal state of an existing game by analyzing the sounds and images it presents, and processing this information, the proposed method makes it possible to create input feedback loops into the game (Fig. 1). Tools such as Picognizer [6] can be used for audio analysis, and ones such as Sikuli [5] can be used to analyze images. This will enable not only the automated operation of games, but also the development of the toolification of games as discussed in Section 5.2, and the insertion of auxiliary inputs as discussed in Section 5.3 to an even higher level of precision.

## 6 Discussion and Future Work

### 6.1 Measures to Address Delay

While the assessment experiments indicated that the delay meets a certain standard, there is room for improvement. Improvements can be made, for example, to the transmission protocol or interface design.

With regard to the former, the DSL4GC is currently expressed in JSON, but the fact that the data size is large indicates that there may be an issue with the utilization of the lightweight MQTT protocol. There is value in compressing the DSL4GC by changing it to binary code, but this comes with a tradeoff in terms of readability.

With regard to the latter, perceived delay can be reduced by regarding the length of the gesture that the user performs as redundant and finishing detection early when detecting and entering a physical gesture. Both present avenues for future research.

### 6.2 Augmenting the Range of Compatible Game Consoles and Devices

The game input emulation element of the system handles keyboards, mouse, and gamepads, but does not cover all the differences among the gaming devices that it works with. The reinforcement of this aspect presents another direction for future research. Besides OS-level S/W emulation and electric-signal-level H/W emulation discussed in this paper, mechanical inputs using robotic devices to existing physical controllers can be another way to implement the game input emulation element [26].

### 6.3 Ethical Issues Concerning Cheating

The objective of this study was to lower the barriers to experimenting with and creating new gaming experiences, and pursuing possibilities for the application of gamification by augmenting possible inputs to games. If the proposed system is used without due consideration, however, it can become a tool to encourage the automated operation of games for one's own benefit. In particular, recent online games, where a user's behavior has an effect on other players, demand prudent use consistent with ethics and rules of the game. Requesting that users agree to engage only in appropriate use or taking system-level measures to prevent abuse in specific games are future challenges in this regard.

1414

## 7 Conclusion

This study proposed middleware, GameControllerizer, that allows users to combine the processes of IoT devices, Web services, and AI, and convert them into game control operations to augment existing digital games. Evidence of its usefulness was provided through an assessment of its performance and a presentation of various use cases. Plans for future work in this area include addressing delays, augmenting the range of compatible game consoles and devices, and taking measures to address cheating while working to increase awareness of the system by hosting such events as hackathons and workshops. Through such efforts, we plan to continue to build a body of work around game design based on GameControllerizer and know-how related to the toolification of games.